\documentclass[journal=nalefd,manuscript=article]{achemso}

\usepackage[version=3]{mhchem} 
\usepackage{mathptmx}
\usepackage{sectsty}
\usepackage{graphicx} 
\usepackage{float}
\usepackage{array}
\usepackage{amsmath}
\usepackage{bm}
\usepackage{xr}

\usepackage{xcolor}
\author{Alessandra Canetta}
\affiliation {\small \textit Institute of Condensed Matter and Nanosciences, Université catholique de Louvain (UCLouvain), 1348 Louvain-la-Neuve, Belgium}

\author{Serhii Volosheniuk}
\affiliation {\small \textit Kavli Institute of Nanoscience, Delft University of Technology, 2628CJ Delft, The Netherlands}

\author{Sayooj Satheesh}
\affiliation {\small \textit Max-Planck-Institut für Festkörperforschung, D-70569 Stuttgart, Germany}

\author{José Pedro Alvarinhas Batista}
\affiliation {\small \textit Nanomat/Q-MAT/CESAM and European Theoretical Spectroscopy Facility, Université de Liège, B-4000, Liège, Belgium}

\author{Alo\"is Castellano}
\affiliation {\small \textit Nanomat/Q-MAT/CESAM and European Theoretical Spectroscopy Facility, Université de Liège, B-4000, Liège, Belgium}

\author{Riccardo Conte}
\affiliation {\small \textit Kavli Institute of Nanoscience, Delft University of Technology, 2628CJ Delft, The Netherlands}

\author{Daniel G. Chica}
\affiliation {\small \textit Department of Chemistry, Columbia University, New York, NY, 10027 USA}

\author{Kenji Watanabe}
\affiliation {\small \textit Research Center for Electronic and Optical Materials, National Institute for Materials Science, 1-1 Namiki, Tsukuba 305-0044, Japan}

\author{Takashi Taniguchi}
\affiliation {\small \textit Research Center for Materials Nanoarchitectonics, National Institute for Materials Science, 1-1 Namiki, Tsukuba 305-0044, Japan}

\author{Xavier Roy}
\affiliation {\small \textit Department of Chemistry, Columbia University, New York, NY, 10027 USA}

\author{Herre S.J. van der Zant}
\affiliation {\small \textit Kavli Institute of Nanoscience, Delft University of Technology, 2628CJ Delft, The Netherlands}

\author{Marko Burghard}
\affiliation {\small \textit Max-Planck-Institut für Festkörperforschung, D-70569 Stuttgart, Germany}

\author{Matthieu J. Verstraete}
\affiliation {\small \textit Nanomat/Q-MAT/CESAM and European Theoretical Spectroscopy Facility, Université de Liège, B-4000, Liège, Belgium}
\alsoaffiliation {\small \textit ITP, Physics Department Utrecht University 3508 TA Utrecht, The Netherlands}

\author{Pascal Gehring}
\email{pascal.gehring@uclouvain.be}
\affiliation {\small \textit Institute of Condensed Matter and Nanosciences, Université catholique de Louvain (UCLouvain), 1348 Louvain-la-Neuve, Belgium}

\title{Impact of spin-entropy on the thermoelectric properties of a 2D magnet}


\begin{document}

\begin{abstract}

Heat-to-charge conversion efficiency of thermoelectric materials is closely linked to the entropy per charge carrier. Thus, magnetic materials are promising building blocks for highly efficient energy harvesters, as their carrier entropy is boosted by a spin degree of freedom. In this work, we investigate how this spin entropy impacts heat-to-charge conversion in A-type antiferromagnet CrSBr. We perform simultaneous measurements of electrical conductance and thermocurrent while changing magnetic order using temperature and magnetic field as tuning parameters. We find a strong enhancement of the thermoelectric power factor around the Néel temperature. We further reveal that the power factor at low temperature can be increased by up to 600\% upon applying a magnetic field. Our results demonstrate that the thermoelectric properties of 2D magnets can be optimized by exploiting the sizeable impact of spin entropy and confirm thermoelectric measurements as a sensitive tool to investigate subtle magnetic phase transitions in low-dimensional magnets.

\end{abstract}

\section{Introduction}

\begin{figure}[t!]
    \centering
    \includegraphics[width=7in]{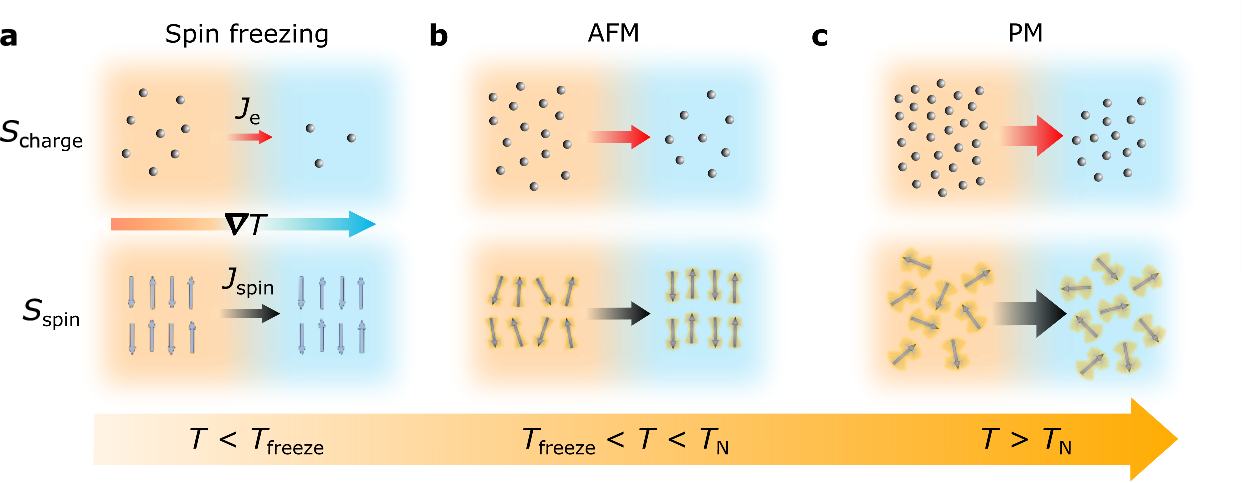}
    \caption{\textbf{Seebeck effect in non-magnetic and magnetic materials.} Schematic illustrating the working principle of the Seebeck effect in a material at different magnetic phases. The three images refer to different temperature ranges and consequently magnetic phases: "spin freezing" state (a), antiferromagnetic state (b) and paramagnetic state (c). The top pictures show the thermally-driven diffusion of the charge carriers, of which direction and magnitude are qualitatively indicated by the red arrows. The bottom images show the additional contribution of the spin entropy ($S_\mathrm{m}$) to the Seebeck coefficient. Direction and magnitude of the entropy flow are qualitatively indicated by the black arrows \cite{Sun21, Yang23}. Temperature ranges are indicated in the large orange arrow at the bottom of the figure, while, in a), the direction of the temperature gradient is illustrated by an orange-to-blue arrow.}
    \label{fig1}
\end{figure}

The Seebeck coefficient ($\alpha$) quantifies the electromotive force or gradient of the electrochemical potential $\boldsymbol{\nabla}V = \boldsymbol{\nabla} \tilde{\mu}/q$ developing in a material exposed to a temperature gradient $\boldsymbol{\nabla} T$ (Fig. \ref{fig1}), and is the central parameter that determines the efficiency of a thermoelectric device \cite{Sun21, Behnia15}. As the electrochemical potential $\tilde{\mu}$ of a population of electrically charged particles consists of the sum of the chemical potential $\mu$ and the electrostatic contribution $q\varphi$, the Seebeck coefficient can be written as \cite{Apertet16}:
\begin{equation}
\alpha = - \frac{\partial \tilde{\mu}}{q\partial T} = - \frac{\partial \mu}{q\partial T} - \frac{\partial \varphi}{\partial T},
\label{eq1}
\end{equation}

where $q$ is the elementary charge. The second term of equation \ref{eq1}, often referred to as effective Seebeck coefficient, contains dynamical effects linked to scattering/carrier relaxation processes \cite{Apertet16, Cai06}. In contrast, the first component -- known as the Kelvin formula \cite{Chaikin76,Peterson10} -- is purely thermodynamic. 
On the basis of thermodynamic considerations for an electronic system, this term is directly related to the average entropy transported per charge carrier\cite{Yang23, Sun21} using the Maxwell equation $\left(\frac{\partial \mu}{\partial T}\right)_{N} = - \left(\frac{\partial S}{\partial N}\right)_{T}$, where $N$ is the mean time-averaged population of the system and $S$ is the electronic entropy \cite{Pyurbeeva22,Yang23,Shastry13}. 
This implies that mechanisms that increase the entropy per carrier can enhance the Seebeck coefficient. In particular, the spin degrees of freedom of carriers in magnetic materials can lead to such increased entropy \cite{Yang23,Tsujii19,Wang23,Portavoce23}. Fig. \ref{fig1} illustrates this concept by comparing the Seebeck effect of an antiferromagnet in three temperature regimes, linked to different magnetic phases. In all cases, under open-circuit conditions, a thermally driven diffusion current of charge carriers (red arrows) from the heated region (depicted in orange) to the cold one (in blue) is balanced by a drift current generated by an electric field that builds up inside the material. Moreover, the so-called \textit{spin-entropy}, $S_\mathrm{m}$, in magnetic materials can contribute to their Seebeck effect (bottom panels of Fig. \ref{fig1}) \cite{Sun21, Yang23}. $S_\mathrm{m}$ is minimum below the "spin freezing" temperature (Fig. \ref{fig1}a, a special magnetic state in CrSBr, see discussion below). Thermal fluctuations will then increase $S_\mathrm{m}$ (Fig. \ref{fig1}b) and it reaches its maximum above the phase transition temperature, as the material enters the paramagnetic state (Fig. \ref{fig1}c) \cite{Sun21}.

In this context, thanks to their controllable magnetism \cite{Yao21, Jiang21, Burch18, Cortie20}, two-dimensional (2D) magnets provide an ideal platform to test this effect. Among the layered van der Waals (vdW) materials A-type antiferromagnet CrSBr stands out for its good cleavability as well as its Néel temperature $T_\mathrm{N}$ of 132 K, one of the highest reported among vdW antiferromagnets \cite{Telford20, Jiang21}. Compared to ferromagnets, AFM materials offer the possibility to change their spin structure into a field-induced FM configuration upon the application of an external magnetic field, adding a degree of freedom in tuning the electronic and thermoelectric properties \cite{Telford22, Bauer2012}.
Each CrSBr van der Waals layer consists of two fused buckled planes of CrS, sandwiched between Br atoms and stacked along the c axis (see Fig. 2a) \cite{Telford20, Jiang21}. CrSBr is an A-type antiferromagnet, with intralayer ferromagnetic (FM) coupling and interlayer antiferromagnetic (AFM) interaction, and with easy/medium/hard axis coinciding with the crystallographic b/a/c axes, respectively \cite{Lee21}. Furthermore, CrSBr shows semiconducting transport properties, with a direct bandgap of $E_\mathrm{G}$ = 1.5 eV and finite electrical conductivity at low temperature\cite{Telford22}. In particular, thanks to the strong coupling between magnetic ordering and transport properties in CrSBr, an external magnetic field can be used to alter the electrical resistance, which tends to decrease as the field increases. This comes as a consequence of the reduction of spin fluctuations, and the different interlayer spin-flip scattering between AFM and FM configurations \cite{Lee21, Telford20, Telford22, Wu22}.
While the electrical transport and magnetic properties of this material have been extensively investigated \cite{Lee21, Telford20, Telford22,Wu22, Lopez22}, the effect of magnetic order on the entropy and thus the thermoelectric properties has not been reported to date.

In this paper we study the impact of electron and spin entropy on the thermoelectric properties of CrSBr thin flakes. To this end, we change magnetic order by varying the sample temperature or by applying an external magnetic field, while simultaneously measuring the electrical and thermoelectric transport properties. We observe a steep increase of the Seebeck coefficient and the thermoelectric power factor with increasing temperature as electrons and spins mobilize, with a local maximum slightly below $T_\mathrm{N}$ which we explain by a competition between electronic band entropy and magnetic entropy in CrSBr. We further reveal that a magnetic field can enhance the power factor by up to 600\% at low temperatures. These findings highlight how (spin-)entropy engineering in 2D magnetic materials could be used to realize thermoelectric heat engines with strongly enhanced performance.

\section{Results}

\begin{figure}[t!]
    \centering
    \includegraphics[width=6.8in]{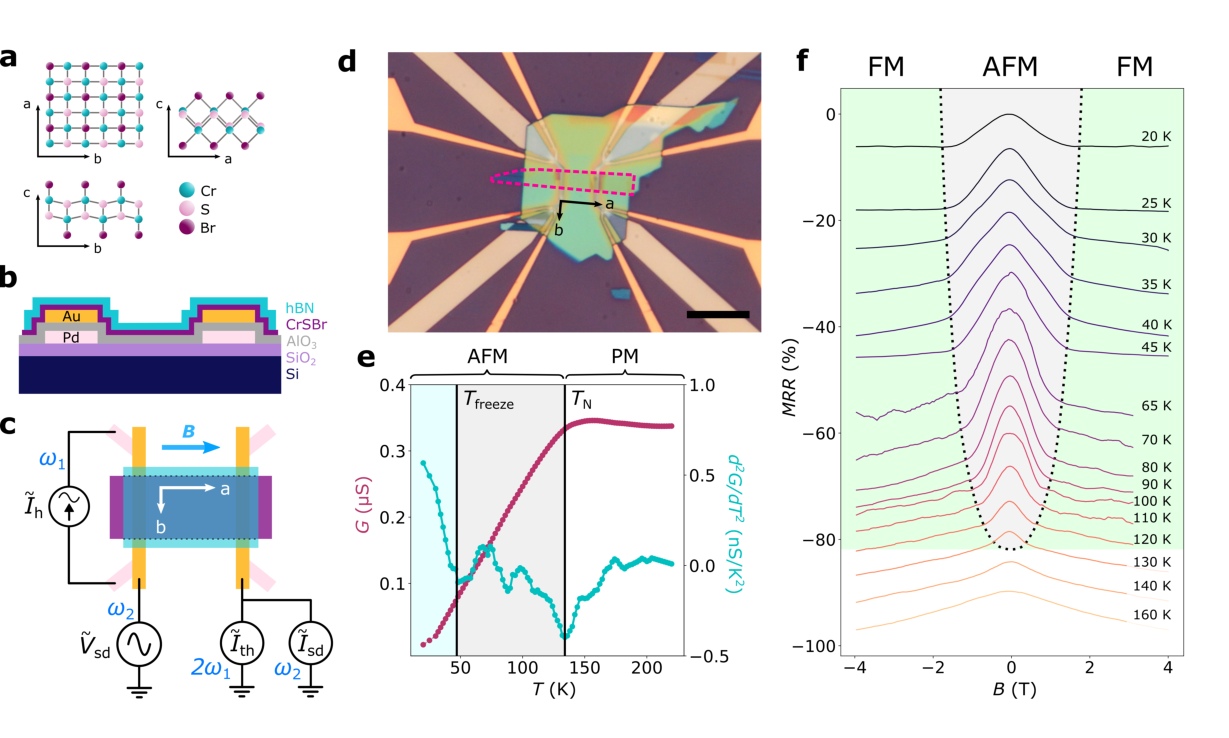}
    \caption{\textbf{Crystal structure, measurement setup, and electrical transport of CrSBr.} a) Crystal structure of CrSBr, from the c axis (top left), a axis (top right) and b axis (bottom). Cr, S and Br atoms are represented as cyan, pink and purple spheres, respectively. b) Side view of the schematic of the device. c) Schematic of the setup used for magnetotransport and thermoelectric measurements. d) Optical image of the measured device. The magenta dashed guideline highlights the position of the CrSBr flake, covered by the hBN layer. e) Temperature dependence of the conductance $G$ (purple) and the second derivative of the conductance (cyan). The white region and the light blue/grey region correspond respectively to the paramagnetic (PM) and antiferromagnetic (AFM) phases of CrSBr. f) Magnetoresistance ratio (MRR) versus the applied magnetic field $B$ at different temperatures between 20 and 160~K. An offset of 6$\%$ is applied for clarity between each pairs of curves. The AFM and FM magnetic phases are shaded grey and green, respectively. The black dotted line defines the saturation field $H_S$ (see Supplementary Information section S10). Scale bar in (d): 10 \textmu m.}
    \label{fig2}
\end{figure}

To measure the electrical and thermoelectric properties of CrSBr thin flakes we employ a device architecture (Fig. \ref{fig2}b-c) that we recently developed for thermoelectric experiments on single molecule junctions \cite{Gehring19,Gehring21}. It consists of pre-patterned contacts, thermometers and microheaters on top of which a CrSBr flake has been stamped using a dry transfer method (see Methods section). A thin hBN flake is used to encapsulate CrSBr to prevent degradation and contamination. An optical micrograph of the final device is shown in Fig. \ref{fig2}d. For a typical measurement (Fig. \ref{fig2}c), an AC current $\tilde{I}_\mathrm{h}$ at frequency $\omega_\mathrm{1}$ is applied to the microheater which generates a temperature bias $\Delta T$ proportional to $\tilde{I}_\mathrm{h}^2$, therefore having frequency 2$\omega_\mathrm{1}$. Simultaneously, an AC voltage $\tilde{V}_\mathrm{sd}$ at frequency $\omega_\mathrm{2} \gg \omega_\mathrm{1}$ is applied to the drain contact. The current to ground on the source contact is then demodulated at frequencies $\omega_\mathrm{2}$ and 2$\omega_\mathrm{1}$ to extract the differential conductance $G=\tilde{I}_\mathrm{sd}/\tilde{V}_\mathrm{sd}$ and the Seebeck coefficient $\alpha =
-\frac{\tilde{V}_\mathrm{th}}{\Delta T} = -\frac{\tilde{I}_\mathrm{th}}{G\Delta T}$ (see Supplementary Information section S9 for details on the temperature calibration), respectively. All magnetic fields in this study were applied parallel to the $a$ (medium) axis of CrSBr.


Fig. \ref{fig2}e illustrates the temperature dependence of $G$ and of $\frac{\mathrm{d}^2G}{\mathrm{d}T^2}$, respectively. $G$ decreases when lowering $T$, typical for semiconducting materials and in good agreement with previous studies \cite{Telford20, Telford22, Wu22}. Furthermore, we observe a maximum in $G$ and a sharp dip in $\frac{\mathrm{d}^2G}{\mathrm{d}T^2}$ around 133 $\pm$ 1 K. We associate this value with the Néel temperature $T_\mathrm{N}$, where the transition from the paramagnetic (PM) state (white region) to antiferromagnetic (AFM) (grey region) occurs \cite{Lopez22,Tsujii19,Telford20,Telford22,Liu22}.
Upon further lowering $T$, $G$ drops by one order of magnitude between $T_\mathrm{N}$ and 20 K \cite{Wu22}. At temperatures lower than  $T_\mathrm{freeze}$ = 47 $\pm$ 2$~K$ the appearance of a low-temperature magnetic hidden order has been reported \cite{Telford22, Lopez22, Boix22}. We do not observe changes in $G$ around $T_\mathrm{freeze}$, however, as we will show later, the Seebeck coefficient changes abruptly below this temperature. 
$G$ values depicted in Fig. \ref{fig2}e are in good agreement with the conductance reported in previous works (\cite{Telford20, Telford22}).


In Fig. \ref{fig2}f we show the magneto-resistance ratio $MRR = \frac{R(B) - R(B=0)}{R(B=0)}\cdot100$ at different temperatures between 20 and 160 K (measurements on additional CrSBr devices can be found in the Supplementary Information section S7). Below $T_\mathrm{N}$, for low magnetic fields, spins are coupled antiferromagnetically between layers and aligned along the $b$ (easy) axis (see grey area). As reported previously, this suppresses the interlayer tunneling, and thus leads to an increase in electrical resistance \cite{Telford20,Telford22}.
By raising the applied magnetic field, spins tend to cant: This re-enables interlayer tunneling and therefore lowers the resistance \cite{Ye22,Telford20,Telford22}. Saturation of the $MRR$ is visible when ferromagnetic order between the layers is established (see green area) \cite{Telford20}.

\begin{figure}[h!]
    \centering\includegraphics[width=6.5in]{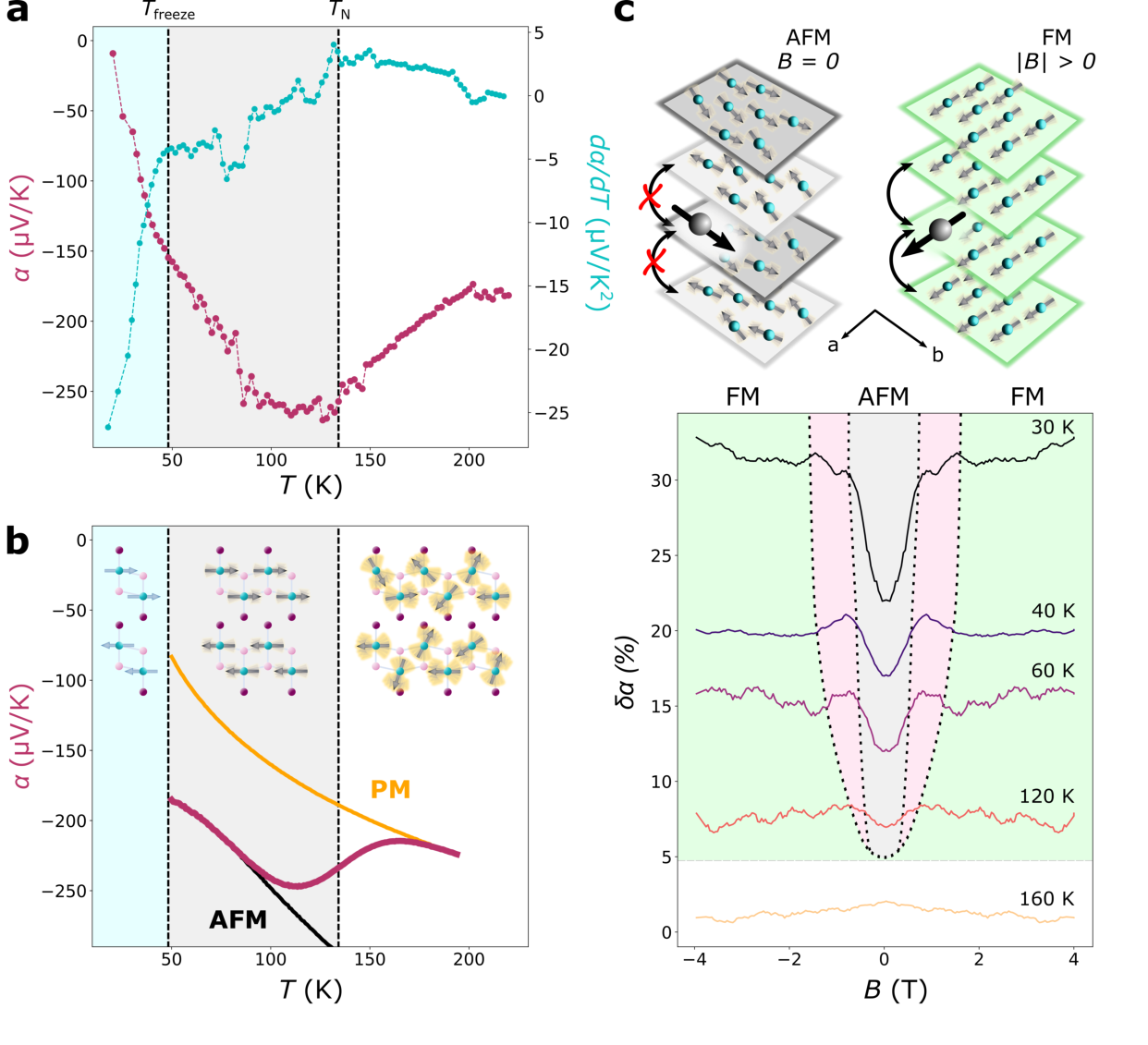}
    \caption{\textbf{Magnetic field and temperature dependence of the Seebeck coefficient of CrSBr.} a) Variation of the Seebeck coefficient (purple curve) and its first derivative (cyan curve) at $B = 0 T$ as a function of temperature. The temperatures $T_\mathrm{N}$ and $T_\mathrm{freeze}$ separate the graph in three areas, colored respectively in white, grey and light blue. b) First principles Seebeck coefficient as a function of T, for a representative (n-type) doping level, calculated in the AFM (black curve), PM (orange curve), and interpolated magnetic states (purple curve). c) Magneto-Seebeck coefficient as a function of temperature. Each curve is offset by 5$\%$ for clarity, and averaged as $\frac{\delta\alpha(B)+\delta\alpha(-B)}{2}$ in order to remove any parasitic effect due to drift in the measurement. The three regions - AFM, transition/canting, FM - are depicted in green, pink and grey, respectively. The field-dependent spin reorientation and interlayer tunneling is illustrated in the top part of the image. The orientation of the crystallographic axes is also reported. An electron residing on one of the layers (dark grey sphere) can tunnel (indicated by the black curved arrows) or not (indicated by the black curved arrows with red X) depending on its spin orientation.}
\label{fig3} 
\end{figure}

Fig. \ref{fig3}a shows the temperature dependence of the Seebeck coefficient simultaneously measured with $G(T)$ (Fig. \ref{fig2}e). The negative sign of $\alpha$ is consistent with the n-type doping typically found in CrSBr, which is attributed to Br vacancies\cite{Telford20, Klein23}. We observe an overall decrease from -265~\textmu V/K to -9~\textmu V/K when cooling the sample from 200~K to 20~K, which is the base temperature of our experiment. Three areas have been highlighted by means of different colors. In the white region ($T > T_\mathrm{N}$), corresponding to the paramagnetic phase, $|\alpha|$ increases as $T$ decreases. $|\alpha|$ reaches its maximum at $T_\mathrm{N}$, stays constant until $T \approx$ 90~K, then decreases (overall about 45\%) until $T_\mathrm{freeze}$ (grey region). When cooling below $T_\mathrm{freeze}$ (light blue area), $|\alpha|$ decreases faster -- as can be seen in the first derivative $\mathrm{d}\alpha / \mathrm{d}T$ (cyan curve) -- down to the value of -9~\textmu V/K at 20~K. To explain this behaviour, we performed first principles calculations within the constant relaxation time approximation \cite{madsen2018_boltztrap2}, for a doping of $\sim 8.10^{18}$ electrons per cm$^3$ (see Supplementary Information Fig. S8). In Fig. \ref{fig3}b we compare the AFM ground state, a collinear PM state (averaging special quasirandom structures \cite{Zunger1990}), and an interpolation between the two \cite{Kormann2014}. The calculations are in good quantitative agreement at low and intermediate temperatures, show the same qualitative extremum and upturn around T$_N$, but underestimate the upward jump of $\alpha$ in the fully PM phase. Our calculation of the bands in a collinear paramagnetic state produces a smaller Seebeck amplitude (less negative). Freeing the spins to be non collinear PM should produce even more phase space and entropy for the spins, and therefore a larger jump. It should be noted that our first principle model is not suitable to predict $\alpha(T)$ at $T<50$~K. In this regime, a strong modulation of the carrier concentration is expected which is not accounted for in the calculations.

To gain further evidence for the impact of magnetic order on the thermoelectric properties of CrSBr, we measured the change in Seebeck coefficient as a function of the applied magnetic field. Fig. \ref{fig3}c shows this magneto-Seebeck coefficient ratio $\mathrm{\delta\alpha}=\frac{\alpha(B) - \alpha(B = 0)}{\alpha(B = 0)}\cdot100$ versus magnetic field $B$ at temperatures varying between 20 K and 160 K. At 160 K, the flake is in a paramagnetic state and the curve shows almost no variation with applied $B$ field. Below $T_\mathrm{N}$ and for small magnetic fields, CrSBr is AFM ordered (grey area) and $\delta\alpha$ is minimum at $B$ = 0 T. As the absolute value of $B$ becomes larger, $\delta\alpha$ increases and reaches a local maximum, then decreases until saturating when FM order is established (green area). The areas including the local maxima of $\delta\alpha$ (in pink) can be identified as transition regions, in which the spins are canting from $a$ to $b$ direction due to the application of external $B$ field \cite{Lopez22}. We observe an increase in $\delta\alpha$ of up to 13\% at low $T$ when changing from AFM to FM order. 

\begin{figure}[t!]
    \centering\includegraphics[width=6.5in]{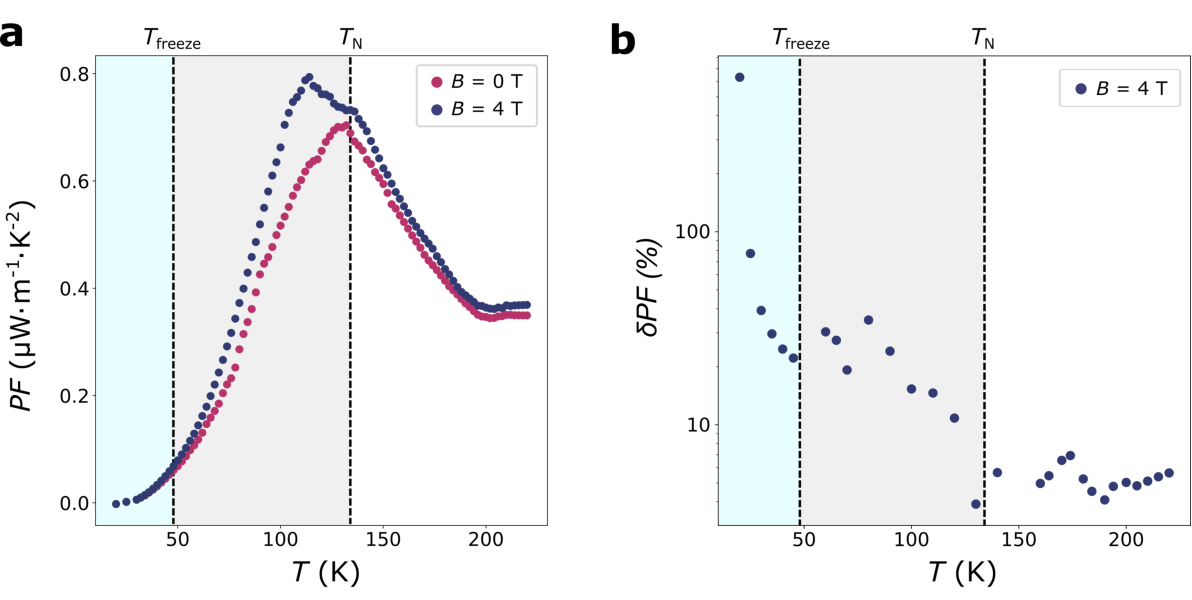}
    \caption{\textbf{Field and temperature dependence of the power factor}. a) Temperature dependent power factor, for $B = 0 T$ (purple) and $B = 4 T$ (blue). b) Magneto-power factor ($\delta PF$) as a function of temperature, measured at $B = 4 T$.}
    \label{fig4}
\end{figure}

Fig. \ref{fig4}a displays the temperature-dependent power factor $PF = \alpha^2 \cdot \sigma$, where $\sigma$ is the electrical conductivity of CrSBr (see Supplementary Information section S3). As part of the figure of merit $zT$, the power factor helps quantifying the energy harvesting efficiency of the material. As it is also proportional to the maximum achievable output power, it is a useful parameter for quantifying Peltier cooling.
At $B$ = 0, $PF$ shows a peak of 7 $\mu$W m$^{-1}$ K$^{-2}$ around $T_\mathrm{N}$, where also the maxima of $G$ and $\alpha$ simultaneously occur. This peak increases in magnitude and shifts to lower temperatures when a magnetic field of $B = 4$~T is applied. Fig. \ref{fig4}b shows the Magneto-power factor $\delta PF = \frac{PF(B = 4\mathrm{T}) - PF(B = 0)}{PF(B = 0) \cdot 100}$ as a function of temperature. We observe that the relative change $\delta PF$ increases with decreasing temperature and reaches values up to 600\% at 20K (see Supplementary Information Fig. S7). Below we will discuss that these findings can be explained by the intrinsic band structure of AFM CrSBr, taking into account variations in the entropy linked to the magnetic order. 

\section{Discussion}

As we described at the beginning of this article, the Seebeck coefficient is closely linked to the entropy $S$ of the system (see Eq. \ref{eq1}) \cite{Pyurbeeva22,Sun22}.
The entropy of a mesoscopic system can be estimated using the Boltzmann formula $S = k_\textrm{B} \ln (\Omega)$, where $\Omega$ represents the number of all possible microstates of the system \cite{Demirel19, Sun21}. 
Here, we assume that $\Omega$ contains three main contributions. $\Omega_\mathrm{p}$ represents the conventional distribution of momenta of the electron gas (electronic band contribution). Then, we take into account a layer degree of freedom $\Omega_\mathrm{layer}$ which quantifies the number of layers a charge carrier can access, as CrSBr is a layered vdW material in which interlayer tunneling is precluded  when switching to AFM order \cite{Wilson21}. Lastly, we include a term $\Omega_\mathrm{s}$ representing all possible spin configurations, which yields the spin-entropy $S_\mathrm{m}$ \cite{Wang03}. The sign of this contribution depends on the nature of the d bands hosting the magnetization, which is positive in CrSBr (hole like, from the d band valence electrons)\cite{Sun21}. Therefore, the electronic and spin entropy contributions have opposite signs.
We now turn back to Fig. \ref{fig3}a, that depicts the temperature dependence of $\alpha$. As $T \geq T_\mathrm{freeze}$, the growth of the Seebeck coefficient abruptly slows down, which is simultaneous with the appearance of a magnetic hidden order below $T_\mathrm{freeze}$. Such hidden order was already observed previously by other groups, who associate its origin either to a magnetic coupling between self-trapped defects \cite{Telford22}, the anisotropic structure of CrSBr -- which can be seen as weakly and incoherently coupled 1D chains \cite{Wu22} --, or a spin-dimensionality crossover caused by a slowing down of the magnetic fluctuations (spin freezing)\cite{Lopez22, Boix22}. The consequence of the spin freezing phenomenon is that spin fluctuations are fully suppressed ($\Omega_\mathrm{s}$ = 1) below $T_\mathrm{freeze}$, and therefore cannot contribute to the entropy \cite{Lopez22} to counteract the electronic $\alpha$ ~\cite{Sun21}. Additionally, interlayer tunneling is suppressed ($\Omega_\mathrm{layer}$ = 1) \cite{Lopez22, Wu22, Boix22}.
As the spins mobilize upon heating, their contribution $S_\mathrm{m}$ is superimposed on the intrinsic electronic $\alpha$. Due to the opposite signs of the electronic and spin entropy contributions, their combined action leads to a plateau and turnover when increasing $T$. At higher temperatures ($T \sim T_\mathrm{N}$), two effects cause the reduction in $|\alpha|$ observed in our experiment: Firstly, fluctuations and $S_\mathrm{m}$ increase as CrSBr approaches $T_\mathrm{N}$, then saturate in the fully paramagnetic phase\cite{Tsujii19, Okabe10}; secondly, an increase in carrier concentration decreases the magnitude of $\alpha$ (less negative, see Supplementary Information Fig. S8). The subsequent increase in $|\alpha|$ beyond temperatures of 200~K, as observed in our experiments and predicted by theory, can be attributed to the dominance of $\Omega_\mathrm{p}$ over the saturated $\Omega_\mathrm{s}$ in the fully paramagnetic (PM) state.

Fig. \ref{fig3}c illustrates how an external magnetic field $B$ affects $\alpha$. The application of $B$ along the $a$ direction of CrSBr produces a continuous canting of the spins \cite{Tsujii19}. Such field-induced spin reorientation initially raises $\Omega_\mathrm{s}$, which leads to an increase in $|\alpha|$ (pink shaded areas). When further increasing $B$, the FM order is established and $\Omega_\mathrm{s}$ is minimized, reducing $|\alpha|$ again \cite{Bonner62, Wang03}. In addition, the transition from AFM to FM order enables interlayer tunneling and thus raises $\Omega_\mathrm{layer}$ (see Fig. \ref{fig3}c). This could explain the higher $\alpha$ in the FM phase compared to the AFM one and has an important consequence: Since both $\sigma$ and $\alpha$ simultaneously increase with magnetic field, the relative change of the power factor ($\alpha^2\sigma$) between AFM and FM order can reach very high values, up to 600\% as observed in our experiment (Fig.~\ref{fig4}b).

\section{Conclusions}

In this work, we investigate the magnetic field and temperature dependent electric and thermoelectric properties of the A-type antiferromagnet CrSBr. We reveal a strong impact of magnetic order on the thermoelectric response of the material, which we attribute to a spin entropy contribution in the total thermopower. In particular, we detect a peak in both the Seebeck coefficient and the power factor around the magnetic transition temperature $T_\mathrm{N}$. 
These findings present a potential way to overcome the limits of conventional thermoelectric devices, by employing magnetic materials. While devices based on CrSBr show enhanced thermoelectric properties at cryogenic temperatures, future research should investigate 2D magnets with higher transition temperature to enable room temperature operation. Promising materials that deserve attention are the recently investigated 2D compounds CrTe\textrm{${_2}$} and Fe\textrm{${_3}$}GaTe\textrm{${_2}$} with magnetic ordering temperatures $>$ 300~K \cite{Wu2021, Zhang2021}. 
To this end, the use of 2D materials adds further benefits, such as the possibility to tune the transition temperature by varying the flake thickness, composition, electrostatic gating, or by producing heterostructures of different layers, in order to yield optimum performance at room temperature \cite{Li12, Razeghi23, Tu20, Oh19}.

\section{Experimental}

\textit{Device fabrication} The thermopower devices have been fabricated by standard Electron Beam Lithography (EBL) on a Silicon (Si) wafer with a 285 nm top layer of Silicon Oxide (SiO\textrm{${_2}$}). First, the heaters are fabricated by depositing 3 nm of Titanium (Ti) and 27 nm of Palladium (Pd). Afterwards, the sample is covered by 10 nm of Aluminum Oxide (Al\textrm{${_2}$}O\textrm{${_3}$}) via atomic layer deposition, performed in an Oxford Instruments FlexAL system. Then, 3 nm of Ti and 47 nm of Gold (Au) are deposited as top contacts.
Crystals of CrSBr were synthesized using a chemical vapor transport method. This synthetic technique involved the transport of material from 950 $^\circ$C at the source side to 850 $^\circ$C at the sink side of a slightly off stoichiometric combination of Cr, S, and CrBr3 in an evacuated fused silica ampoule. The detailed synthesis and cleaning procedure can be found elsewhere \cite{Scheie22}. CrSBr flakes are mechanically exfoliated using the Scotch tape method \cite{Novoselov04,Novoselov05} and deposited on a PDMS square of approximately 1 mm x 1 mm positioned on a glass slide to facilitate its handling \cite{Castellanos13, Frisenda17}, and transferred onto the pre-patterned contacts. Encapsulation with hBN is then performed by means of the dry transfer method using a PDMS-Polypropylene carbonate (PPC) stamp \cite{Kinoshita19, Wang13}.
The thickness of all flakes is identified by optical contrast and then confirmed by atomic force microscopy measurements (see Supplementary Information section S2). CrSBr handling is entirely done under inert atmosphere in a N\textrm{${_2}$} glove box, with $<$0.5 ppm of O\textrm{${_2}$} and $<$0.5 ppm of H\textrm{${_2}$}O content, in order to avoid air degradation and contamination of the sample. More details about the fabrication process can be found in the Supplementary Information (see sections 1-4-5).

\textit{Thermoelectric and electrical transport measurements} 
Electrical transport and thermopower measurements were performed in a $^4$He cryostat using home-built ultra-low noise voltage/current sources and pre-amplifiers. We employed a lock-in double-demodulation technique\cite{Gehring21}, which allows to decouple the thermocurrent $\tilde{I}_\mathrm{th}$ flowing as a response to a thermal bias $\Delta T$ from $\tilde{I}_\mathrm{sd}$ , the response to a voltage bias $\tilde{V}_\mathrm{sd}$ \cite{Gehring21}. To this end, an AC current $\tilde{I}_\mathrm{h}$ = 0.5~mA (power $P$ = 0.12 mW) at frequency $\omega_\mathrm{1}$ = 3~Hz is applied to the Pd heater with a Stanford Research SR830 lock-in connected to a current source. Simultaneously, an AC voltage $\tilde{V}_\mathrm{sd} = 10$~mV at $\omega_\mathrm{2}$ = 13~Hz is applied to the drain contact. The current at the source contact is pre-amplified by a low-noise transimpedance amplifier and demodulated at $\omega_\mathrm{1}$ and $\omega_\mathrm{2}$ to obtain $\tilde{I}_\mathrm{th}$ and $\tilde{I}_\mathrm{sd}$, respectively (see Supplementary Information section 6).

\section{Theory}

\textit{First Principles Simulations} 
We employ Density Functional Theory\cite{Hohenberg_1964_DFT,kohn_1965_DFT_LDA} as implemented in the ABINIT\cite{gonze_2020_abinit} software suite, in a projector augmented wave\cite{blochl_1994_PAW} basis using JTH\cite{Jollet_2014} format atomic datasets with plane wave kinetic energy cutoffs of 25 and 30 Ha for the wavefunctions and density (total energies are converged to within 1 meV/atom). The first Brillouin Zone is sampled using a uniform grid of 13$\times$ 9$\times$ 3 points for the ground state and 26$\times$ 18$\times$ 6 to prepare transport calculations. 
The exchange correlation functional used was the generalized gradient approximation of Perdew Burke and Ernzerhof\cite{perdew_1996_PBE_GGA}, augmented by the Grimme D3 van der Waals dispersion\cite{Grimme_2010_D3}.
The magnetism of Cr necessitates additional Hubbard repulsion in the DFT+U method\cite{anisimov_1991_ldaplusU} (Fully Localized Limit) with a U of 4 eV and J of 1 eV on Cr only (effective U of 3 eV). Most calculations are performed for collinear spins, and additional checks with non collinear magnetization and spin orbit coupling. The electronic bands are occupied with a Gaussian smearing of 10$^{-4}$ Ha to improve convergence and allow for variable spin polarized occupations, though the final band structures are all semiconducting.
The PM high temperature phase is approximated using the Special Quasirandom Structure (SQS) approach\cite{Zunger1990} implemented in the icet package\cite{ngqvist2019} to generate spin configurations in the infinite temperature limit\cite{Abrikosov2016}. By averaging over 5 different $3\times3\times2$ supercells with these disordered spins configurations, we obtain a representative electronic DOS and states for the high temperature phase.

\textit{Transport}
The transport coefficients are calculated within the constant relaxation time approximation using the Boltztrap2 code\cite{madsen2018_boltztrap2} for constant doping levels. The Seebeck coefficient is obtained quantitatively as the (unknown) relaxation time drops out of its expression. We infer the experimental doping level ($\sim 8 \cdot 10^{18}$ carriers per cm$^3$) by comparison of S(T) at low temperature, and consider it constant above $T_\mathrm{N}$ as well.

We apply the Boltztrap2 code to the PM approximant SQS supercells, then the results are averaged, to estimate the transport of the PM phase. To compare to the experimental transport measures, we add a crossover/switching from AFM to PM, inspired by Ref.~\cite{Kormann2014}. The switching is chosen as an error function (erf) centered at $T_\mathrm{N}$ with a width of 30 K.

\section*{Author Contributions}
P.G. conceived and supervised the experiments. S.V. fabricated the pre-patterned contacts and performed the experiments. A.C. prepared the CrSBr device, performed the experiments with the support of S.S. and M.B., and evaluated the data. A.C. and P.G. wrote the manuscript. D.C. and X.R. synthetized the CrSBr. K.W. and T.T. provided the hBN. A.Castellano, J.A-B, and M.V. performed the first principles simulations. All authors have given approval to the final version of the manuscript.

\section*{Conflicts of interest}
There are no conflicts to declare.

\begin{acknowledgement}
P.G. acknowledges financial support from the F.R.S.-FNRS of Belgium (FNRS-CQ-1.C044.21-SMARD, FNRS-CDR-J.0068.21-SMARD, FNRS-MIS-F.4523.22-TopoBrain), and from the EU (ERC-StG-10104144-MOUNTAIN).
P.G, A.Canetta, A.Castellano, J.A-B, and M.V. acknowledge funding from the Federation Wallonie-Bruxelles through the ARC Grant No. 21/26-116, and from the FWO and FRS-FNRS under the Excellence of Science (EOS) programme (40007563-CONNECT). H.v.d.Z acknowledges support by the FET open project QuIET (Number 767187) and by the Netherlands Organisation for Scientific Research (NWO). 
K.W. and T.T. acknowledge support from the JSPS KAKENHI (Grant Numbers 21H05233 and 23H02052) and World Premier International Research Center Initiative (WPI), MEXT, Japan for the growth of h-BN crystals. M.B. is grateful for support from the Deutsche Forschungsgemeinschaft (DFG) via Grant BU 1125/11-1.
M.V. acknowledges a PRACE award granting access to MareNostrum4 at Barcelona Supercomputing Center (BSC), Spain and Discoverer in SofiaTech, Bulgaria (OptoSpin project id. 2020225411).
Synthetic work at Columbia was supported by the National Science Foundation (NSF) through the Columbia Materials Science and Engineering Research Center on Precision-Assembled Quantum Materials (DMR-2011738).

\end{acknowledgement}

\bibliography{biblio}

\end{document}